\documentclass[fleqn,usenatbib]{mnras}

\usepackage{newtxtext,newtxmath}

\usepackage[T1]{fontenc}
\usepackage{subfig}
\DeclareRobustCommand{\VAN}[3]{#2}
\let\VANthebibliography\thebibliography
\def\thebibliography{\DeclareRobustCommand{\VAN}[3]{##3}\VANthebibliography}

\usepackage{caption}
\usepackage{graphicx}
\usepackage{amsmath}	
\usepackage{nccmath}
\usepackage[mode=text]{siunitx}
\usepackage{comment}
\usepackage{ulem}

\title[]{Formation of $\text{H}_{2}$ on polycyclic aromatic hydrocarbons under conditions of the ISM: an ab initio molecular dynamics study}

\author[NF-Barrera et al.]{
Nicol\'{a}s F. Barrera,$^{1}$
Patricio Fuentealba,$^{1,2}$\thanks{E-mail: pfuentea@hotmail.es (PF); cacarden@gmail.com (CC)}
Francisco Mu\~noz,$^{1,2}$
Tatiana G\'omez,$^{3}$
Carlos C\'{a}rdenas$^{1,2}$\textcolor{blue}{\footnotemark[1]}
\\
$^{1}$Universidad de Chile, Facultad de Ciencias, Departamento de F\'{i}sica, Av. Las Palmeras 3425, \~{N}u\~{n}oa, Santiago, Chile\\
$^{2}$Centro para el Desarrollo de la Nanociencia y la Nanotecnolog\'{i}a (CEDENNA), Av. Ecuador 3493, Santiago 9170124, Chile\\
$^{3}$Theoretical and Computational Chemistry Center, Institute of Applied Chemical Sciences, Faculty of Engineering, Universidad Aut\'{o}noma de Chile, Santiago, Chile\\
}

\date{Accepted XXX. Received YYY; in original form ZZZ}

\pubyear{2023}

\begin{document}
\label{firstpage}
\pagerange{\pageref{firstpage}--\pageref{lastpage}}
\maketitle

\begin{abstract}
Understanding how the $\mathrm{H}_2$ molecule is formed under the chemical conditions of the interstellar media (ISM) is critical to the whole chemistry of it. Formation of $\mathrm{H}_2$ in the ISM requires a third body acting as a reservoir of energy.  Polycyclic aromatic hydrocarbons (PAHs) are excellent candidates to play that role. In this work we simulated the collisions of hydrogen atoms with coronene  to form  $\mathrm{H}_2$ via  the  Eley-Rideal mechanism. To do so, we used Born-Oppenheimer (\textit{ab initio}) Molecular Dynamics simulations. Our results show that that adsorption of  H atoms and subsequent release  of $\mathrm{H}_2$ readily happen on coronene for H atoms with kinetic energy as large as 1 eV. Special attention is paid to dissipation and partition of the energy released in the reactions.  The capacity of coronene to dissipate  collision and reaction energies depends varies with the reaction site. Inner sites dissipate energy easier and faster than edge sites, thus evidencing an interplay between the potential energy surface around the reaction center and its ability to cool the projectile. As for the the recombination of H atoms and the subsequent formation of $\mathrm{H}_{2}$, it is observed that $\sim 15\%$ of the energy is dissipated by the coronene molecule as vibrational energy and the remaining energy is carried by $\mathrm{H}_{2}$. The $\mathrm{H}_{2}$ molecules desorb from coronene with an excited vibrational state ($\upsilon \geq 3$), a large amount of translational kinetic energy ($\geq$ 0.4 eV) and with a small activation of the rotational degree of freedom.

\end{abstract}

\begin{keywords}
astrochemistry -- molecular processes -- ISM: molecules.
\end{keywords}



\section{Introduction}

Molecular hydrogen is the most abundant molecule in the Universe \citep*{wakelam2017h2, vidali2013h2}. Hence, $\mathrm{H}_2$ plays a key role in the physical and chemical properties of the interstellar media (ISM) as it acts as a cooling/heating agent,  triggers the collapse of molecular clouds which lead to star formation, and it is  a precursor in  networks of chemical reactions in interstellar dust and molecular clouds \citep*{le2009molecular, barrales1}. For example, $\mathrm{H}_{2}$ facilitates the formation of molecules and ions such as $\mathrm{CO}, \mathrm{H_{2}O}, \mathrm{HCN}, \mathrm{HCO}^{+}$ and $\mathrm{H}_{3}^{+}$ \citep{boschman2015h2}.  The mechanisms by which $\mathrm{H}_2$ could be formed in the conditions of the ISM are not completely understood. What is clear is that abundance of $\mathrm{H}_{2}$ in the ISM can not be explained as a the direct association of two H atoms in gas phase. \citep{wakelam2017h2, vidali2013h2, mennella2011catalytic}. $\mathrm{H}_{2}$ formation is a very exothermic process and the  $\sim 4.5$\,eV released in the reaction can not be effectively dissipated by the particle density of the ISM. The  consensus is that $\mathrm{H}_{2}$ formation takes place mainly on dust grains, which act as reservoir  that  can absorb part of the  energy released in the association reaction \citep{van1948evolution, Gould1963,lemaire2010,pantaleone2021}. There is evidence that, under conditions relevant to the ISM, $\mathrm{H}_{2}$ forms on surfaces of silicates \citep{Pirronello1997,lemaire2010}, amorphous water \citep{Roser2003}, graphite, carbonaceous material \citep*{latimer2008}, and polycyclic aromatic hydrocarbons (PAHs) \citep{bauschlicher1998,thrower2012,Roser2003}.\\

Interstellar dust grain surfaces have several  mechanisms for $\mathrm{H}_{2}$ formation. Two of primary importance are  the Langmuir-Hinshelwood (LH) and the Eley-Rideal (ER) mechanisms \citep{bauschlicher1998,mennella2011catalytic}. The LH mechanism involves atoms being absorbed on the surface from the gas phase, diffusing and then reacting. Depending on how the energy released in the reaction is accommodated by the host grain, $\mathrm{H}_{2}$ molecules might or might not leave the surface. In contrast, the ER mechanism involves chemisorption of an H atom on the surface and a second H atom impacting in the cross-section of the first to abstract it, forming $\mathrm{H}_{2}$ by recombination. The nascent $\mathrm{H}_{2}$ molecule is likely to leave the surface with much of the energy released in the reaction \citep{wakelam2017h2}. Both mechanisms require hydrogen atoms to absorb on the surface of the dust grain. As the grain temperature increases, physisorbed H atoms start to evaporate, and only chemisorbed H atoms remain on the dust grain surface. Therefore, the LH mechanism becomes less efficient, and $\mathrm{H}_{2}$ formation relies on the ER mechanism. Physisorbed H atoms desorb at grain temperatures between $10-20$\,K \citep*{Cazaux2009}.\\

Although the formation of molecular hydrogen  via dust grains in the ISM is widely accepted, there are still gaps in our knowledge, particularly in high-kinetic-energy/temperature conditions and on surfaces different  than ice water \citep{wakelam2017h2, vidali2013h2}. Some researchers proposed that polycyclic aromatic hydrocarbons may play a key role in $\mathrm{H}_{2}$ formation. PAHs are components of interstellar dust grains found in various environments throughout the ISM \citep*{tielens2008interstellar, burke2010ice}. They are estimated to account for 2 to 30 percent of the carbon in the ISM \citep*{tielens2008interstellar} and amount up to half of the total dust grain surface area available for chemical reactions \citep*{Weingartner2001}. Recently, researchers detected PAH molecules in the TMC-1 molecular cloud, including 1- and 2-cyanonaphthalene \citep{mcguire2021detection, mccarthy2021aromatics} and indene \citep{Cernicharo_2021}. Previous experimental and theoretical studies have stablished the potential of PAHs as catalysts for $\mathrm{H}_{2}$ formation \citep{thrower2012,foley2018,boschman2015h2,barrales2}. For instance, a prototypical PAH molecule, pyrene, has a barrierless reaction channel for $\mathrm{H}_{2}$ recombination in the singlet state (ground state), but a barrier in the triplet state  \citep{barrales1}.\\

We examine the dynamics of chemisorption of H atoms and the formation of $\mathrm{H}_{2}$ on coronene, a prototype PAH, under conditions relevant to the ISM and adopting the ER reaction mechanism. Since these reactions release significant amounts of energy \citep{cortes2014, wakelam2017h2} ($\Delta E \leq -0.6$\,eV), we are interested in understanding coronene's ability to absorb the energy generated by both reactions under different impact parameters. We also aim to gain insight into the distribution of kinetic energy of the nascent $\mathrm{H}_{2}$ molecule, which other authors have referred to as the fate of the energy released in the formation of $\mathrm{H}_{2}$ \citep{lemaire2010,pantaleone2021}. We employed Born-Oppenheimer molecular dynamics (BOMD) of the impact of $\mathrm{H}\cdot$ projectiles on coronene ($\mathrm{C}_{24}\mathrm{H}_{12}$). Previously, BOMD has been used to study the dissipation of the energy released in $\mathrm{H}_{2}$ formation \citep{pantaleone2021}, and also coronene has been used in the past as a model of $\mathrm{H}_{2}$ formation on cations of PAH \citep{foley2018}. The remainder of this article is organized as follows. In Section 2 we describe the computational and chemical model. The  results of our simulations and its implications are presented and discussed in Section 3. We close with conclusions in Section 4.

\section{Computational Details}
\begin{figure}
\centering
	\includegraphics[scale=0.25]{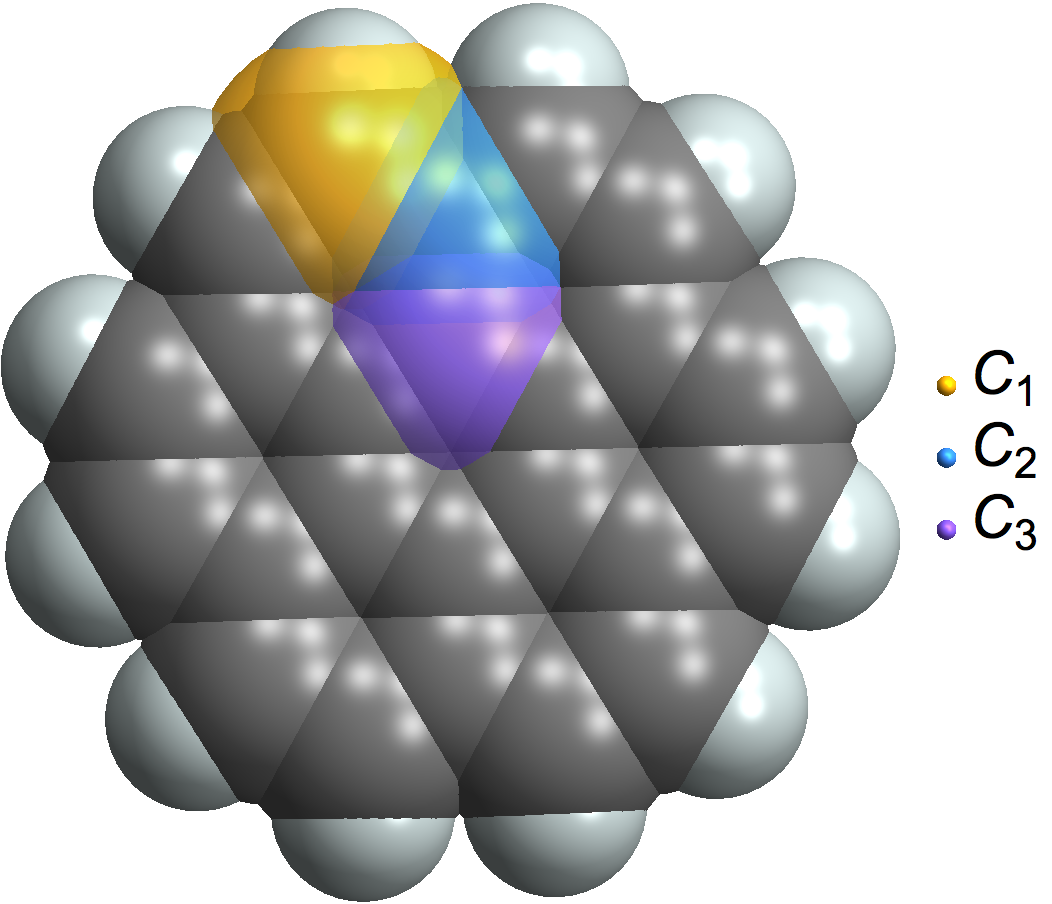}
    \caption{Labels of the differents adsorption sites on coronene. The colour show the equivalent carbons in a particular ring.}
    \label{fig:Cor_positions}
\end{figure}

We used a coronene molecule as a prototype of PAH in collision with $\mathrm{H}$ atoms. We assume that the coronene molecule is in the equilibrium geometry of its electronic singlet ground state (see Figure \ref{fig:Cor_positions}). We used BOMD in the microcanonical ensemble, i.e., at constant energy and number of atoms. This ensemble fits best the conditions of the ISM where the low concentration of particles hinders the dissipation of thermal energy on timescales of the collision (< 500 fs). Similar setups have been used by some us \citep{Inostroza_2019} and others \citep{Inostroza_2020, Inostroza_2021, Chen_2021,pantaleone2021} to simulate molecular collisions in the ISM.\\

Electronic states, energies and forces on atoms were calculated using density-functional tight-binding method (SCC-DFTB or also called DFTB2), a parameterized quantum chemical method derived from density functional theory (DFT) based on a second-order expansion of the DFT total energy around a reference density \citep{Foulkes1989,Koskinen2009,Elstner2014}. In particular, for the first step of ER mechanism (chemisorption) we used the spin-polarized DFTB2 with MIO Slater–Koster parameter set \citep{Elstner1998,Kohler2005}. All theses calculations were carried out using DFTB+ package \citep{DFTB+}. For the second step (abstraction) we used the uDFTB2 with MIO set of parameters augmented with  MIOmod parameters for the H-H potential \citep{Elstner1998}. We chose MIOmod H-H parameters over the MIO because the former allows a better description of the atomization energy, bond length and vibrational frequency of the  $\mathrm{H}_{2}$ molecule. All theses calculations were done with  Gaussian 09 \citep{frisch2009gaussian}. We tested the suitability of the tight-binding parametrization by comparing the potential energy of randomly-selected trajectories with DFT calculation employing PBE functional \citep*{PBE}  and the same basis set, 6-31G(d), as the DFTB parametrizations (see Section 2 in Supporting Information). Finally, we used the Verlet-velocity algorithm to integrate the equations of motion. A assessment of the conservation of energy indicates that a time step of $0.2$\,fs  guarantees conservation within a 50 meV interval during the whole simulation. \\

\begin{figure}
\centering
	\includegraphics[scale=0.40]{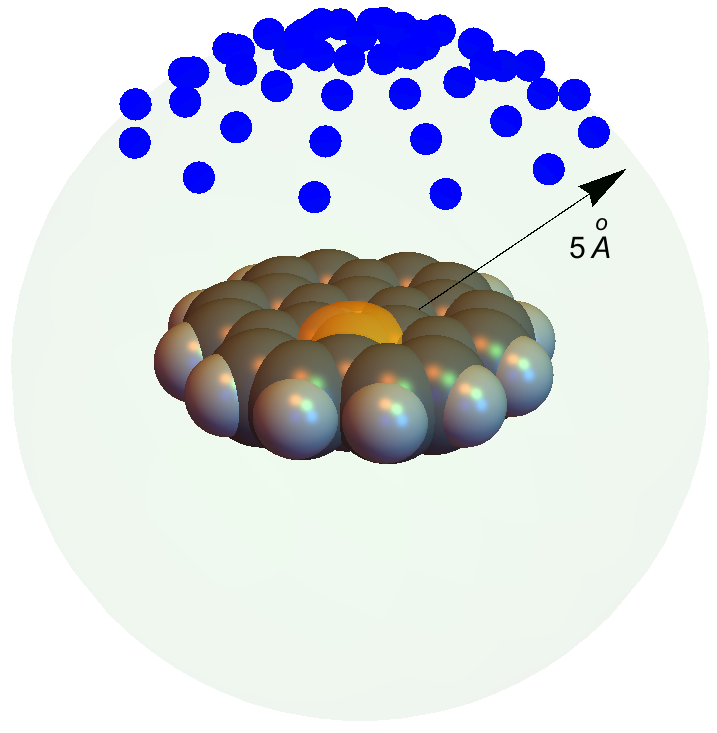} 
    \caption{Initial position of H atoms used for the molecular dynamics of the collision of $\mathrm{H}\cdot$ and coronene. Each blue point on the surface represents an initial position of the $\mathrm{H}\cdot$ projectile. The 49 points (blue) are distributed over the surface of the spherical cap of radius  $\SI{5}{\angstrom}$ and centered in the target atom (orange), which could be be a carbon (for chemisorption) or and adsorbed H (for $\mathrm{H}_2$ abstraction).}
    \label{fig:Coroneno_orienta}
\end{figure}

As coronene belongs to  $D_{6h}$ point group, it has three nonequivalent carbons, which we labeled  $C_1$, $C_2$ and $C_3$ in Figure \ref{fig:Cor_positions}. For each site  49 projectiles were launched from an spherical cap of radius $\SI{5}{\angstrom}$ centered in the target carbon (see Figure \ref{fig:Coroneno_orienta}). This distance prevents energy transfer (interactions) between the surface and the projectile at the beginning of the molecular dynamics. The cap was discretized with the polar angle ranging  from $0^{\circ}$ to $40^{\circ}$ with an step of  $10^{\circ}$, while the azimuthal angle ranges from $0^{\circ}$ to $360^{\circ}$ with $30^{\circ}$ step. Three values of initial kinetic energy of the H projectile were used for each carbon site: (\textit{i}) a value slightly higher than the reaction barrier, (\textit{ii}) 0.50, and (\textit{iii}) 1.00 eV. \\

\section{RESULTS AND DISCUSSION}
\subsection{Chemisorption of a single hydrogen atom}

In the ER mechanism, the initial step involves the chemisorption of the incoming hydrogen atom on any of the three available sites, which have different properties. The reactions are exothermic and have a reaction barriers, see Table \ref{tab:impact_energies}. The smallest  selected kinetic energy of the projectiles is slightly higher than the reaction barrier to allow  trajectories where the  reaction occurs. The regime of medium (0.5 eV) and large (1.0 eV) kinetic energy is also explored. We analyzed the energetics of the dynamics on two steps. First, we examined how likely it is for $\mathrm{H}\cdot$ projectiles to stick to the molecule. Second, we investigated how much energy the reactive site retains, and how quickly this energy dissipates to the rest of the molecule.\\

\begin{figure*}
\centering
	\includegraphics[scale=0.5]{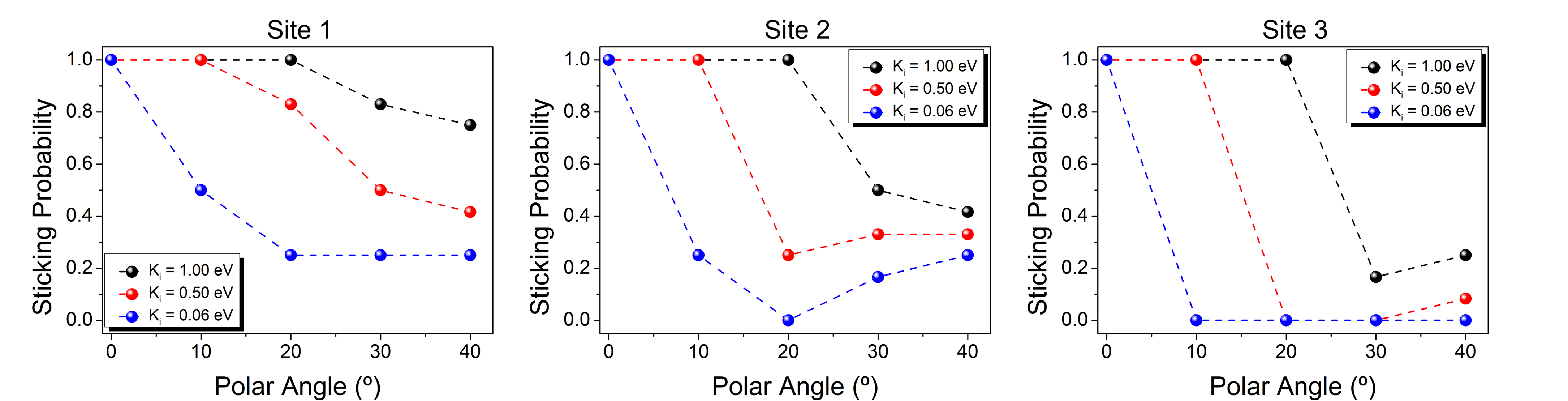}
    \caption{Sticking probability for chemisorption of H on coronene.}
    \label{fig:Sticking1}
\end{figure*}

The hydrogen atom hitting the coronene with kinetic energy higher than the minimum reaction barrier doesn't always result in H chemisorption. This depends on the orientation of the projectile and its kinetic energy. When the kinetic energy is only marginally greater than the barrier, collisions might not be reactive because the impact channel does not coincide with the minimum energy path, which is perpendicular collision (polar angle = $0^{\circ}$). This is clear from Figure~\ref{fig:Sticking1} where the  sticking coefficient (the fraction of incoming H that adsorb on coronene) is plotted as a function of the projectile kinetic energy and polar angle. At the lowest projectile kinetic energy the chance of H to adsorb greatly depends on the polar angle. This dependence, however, does not follow the dependence expected in thermal equilibrium (\textit{i.e.} $\propto \cos^2(\theta) $).  The general trend is that sticking probability decreases as the projectile deviates from normal impact and it is completely suppressed for slant impacts on the innermost carbon ($C_3$). As the projectile kinetic energy increases to 0.5 and 1.0 eV, the sticking probability for non-normal impacts markedly increases. For instance, for a projectile with  1 eV of kinetic energy at least 70~\% of impacts at low polar angle ($40^{\circ}$) result in chemical adsorption of the H on the carbon  atoms on the edge of the molecule ($C_3$). Low-angle impacts on inner carbon atoms again are less likely to result in adsorption but the sticking probability increases with the projectile energy. In this case, over 20\% of collisions on $C_3$  are reactive  when the H atoms carries 1.0 eV. Sticking probabilities show that the PAH efficiently accommodates the reaction and impact energies. The chemical adsorption of H depends more on how the topology of the potential energy surface around carbon hinders the approach of hydrogen to the surface.\\

\begin{figure*}
    \centering
    \includegraphics[scale=0.5]{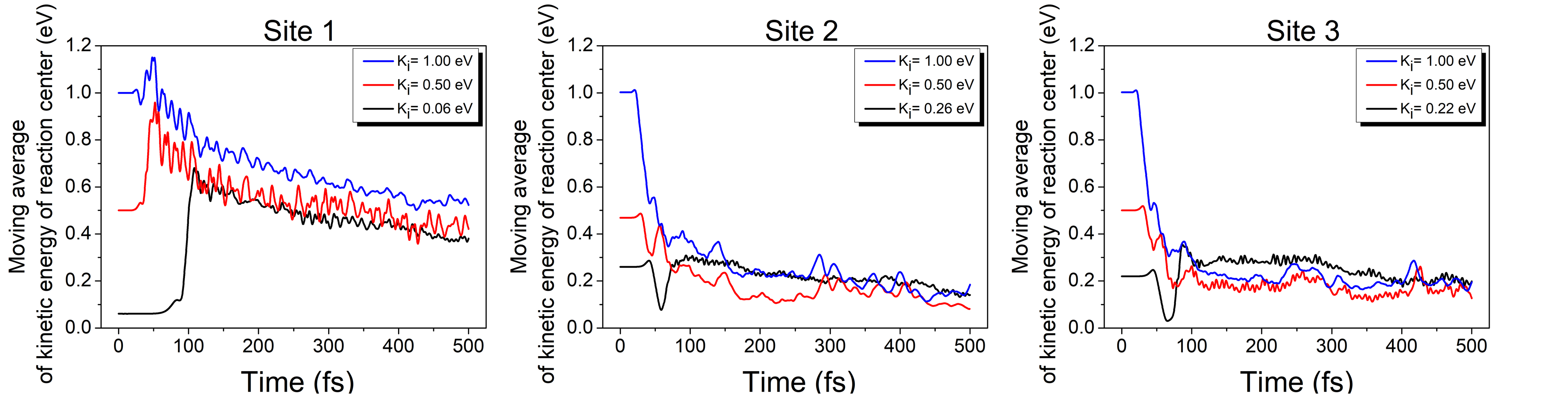}
    \caption{Moving average of kinetic energy of the reactive site for the three nonequivalent sites ( $C_1,C_2$ and $C_3$) of the coronene.}
    \label{fig:MV_Kin_Ener_RS}
\end{figure*}

To better understand how the PAH absorbs collision energy and energy from the exothermic adsorption of hydrogen, we analyzed the average kinetic energy of the incoming hydrogen and the reaction center that is impacted, which includes the target carbon and any bonded hydrogen. This sum is referred to as the kinetic energy of the reaction center, and we plotted its changes over time in Figure~\ref{fig:MV_Kin_Ener_RS}. At the start of the collision, this energy is equal to the projectile's kinetic energy, as the PAH has none. The kinetic energy of the reaction center decreases quickly over time for all impact sites and energies, demonstrating the molecule's ability to absorb excess energy for chemisorption of the hydrogen. After 500 fs, the reaction centers in the molecule's inner regions ($C_2$ and $C_3$) have cooled to around 0.2 eV, a third the energy required to desorb the hydrogen (refer to $\Delta E^{\circ}$ in Table \ref{tab:impact_energies}). This warranties that once H is chemisorbed on a C atom, it remains bonded. The reaction center at the molecule's edge ($C_{1}$), however, doesn't cool as much as the inner centers and maintains around 0.5 eV of kinetic energy after 500 fs. The reason for this is that carbon atoms at the edge of the molecule have fewer neighbors carbon atoms to couple vibrationally and dissipate energy with. Overall, there is an interplay between the topology of the potential energy surface (PES) around a given site and its ability to cool the projectile: at the edge it sticks easier at low polar angle, but is less efficient at accommodating energy.

\begin{table}
\centering
	\caption{Activation energy $\Delta E^{\ddagger}$, reaction energy $\Delta E^{\circ}$  and impact energies $\mathrm{K}_{i}$ of hydrogen projectiles for each site. All values in eV.}
	\label{tab:impact_energies}
	\begin{tabular}{cccc} 
		\hline\hline
		Site $C_{i} (i = 1-3)$ &  $\Delta E^{\ddagger}$ & $\Delta E^{\circ}$ & Energy of projectiles $\mathrm{K}$ (eV)\\

		\hline\hline
		1 & 0.060 & -1.45 & 0.06, 0.5, 1.0 \\
		2 &  0.220 & -0.74 & 0.26, 0.5, 1.0 \\
		3 & 0.196 &  -0.63  & 0.22, 0.5, 1.0 \\
		\hline\hline
	\end{tabular}
\end{table}
\subsection{$\mathrm{H}_{2}$ formation via Eley-Rideal mechanism}

\begin{figure*}
\centering
\includegraphics[scale=0.5]{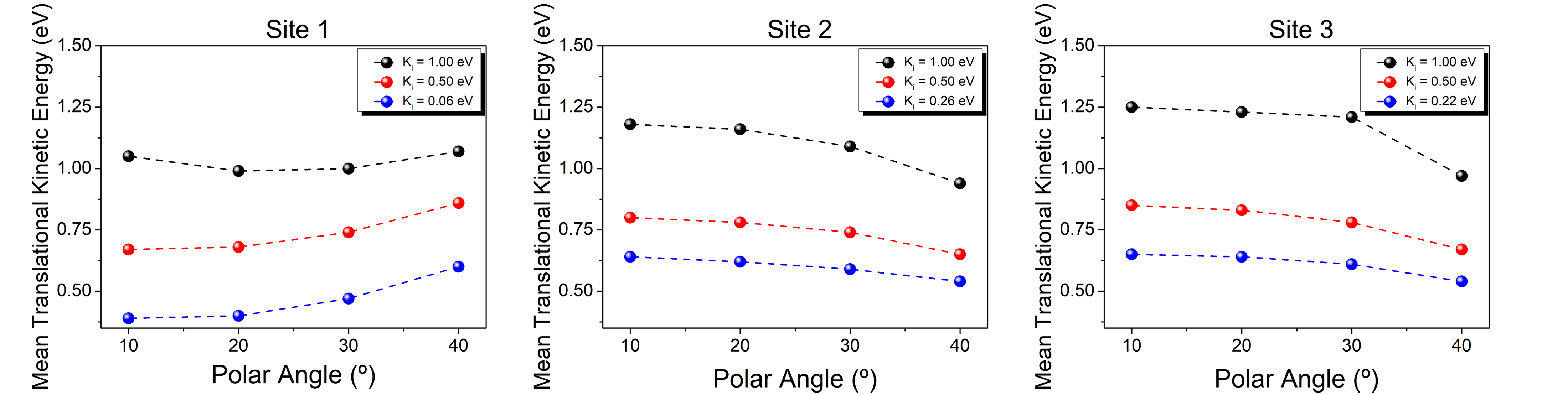}
    \caption{Mean translational kinetic energy of the $\mathrm{H}_{2}$ molecules averaged over the azimuthal angle of the incoming $\mathrm{H}$ projectile for each site: $C_1,C_2$ and $C_3$.}
    \label{fig:MTKE}
\end{figure*}

\begin{figure*}
    \includegraphics[width=2.0\columnwidth]{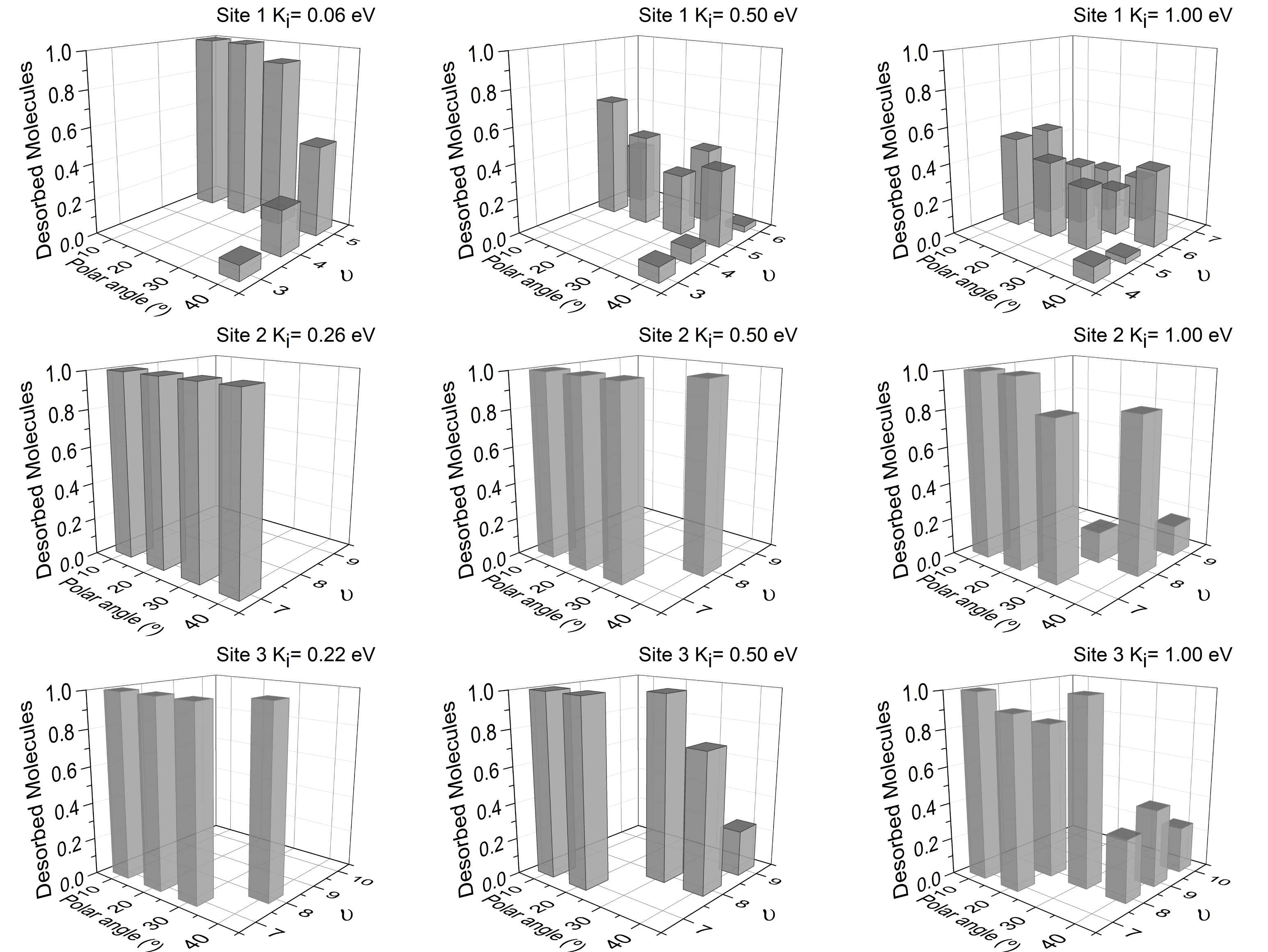}
 \caption{Vibrational population for the $\mathrm{H}_{2}$ nascent onto the three sites.}
 \label{f:H2_v}
\end{figure*}

\begin{figure*}
\centering
\includegraphics[scale=0.20]{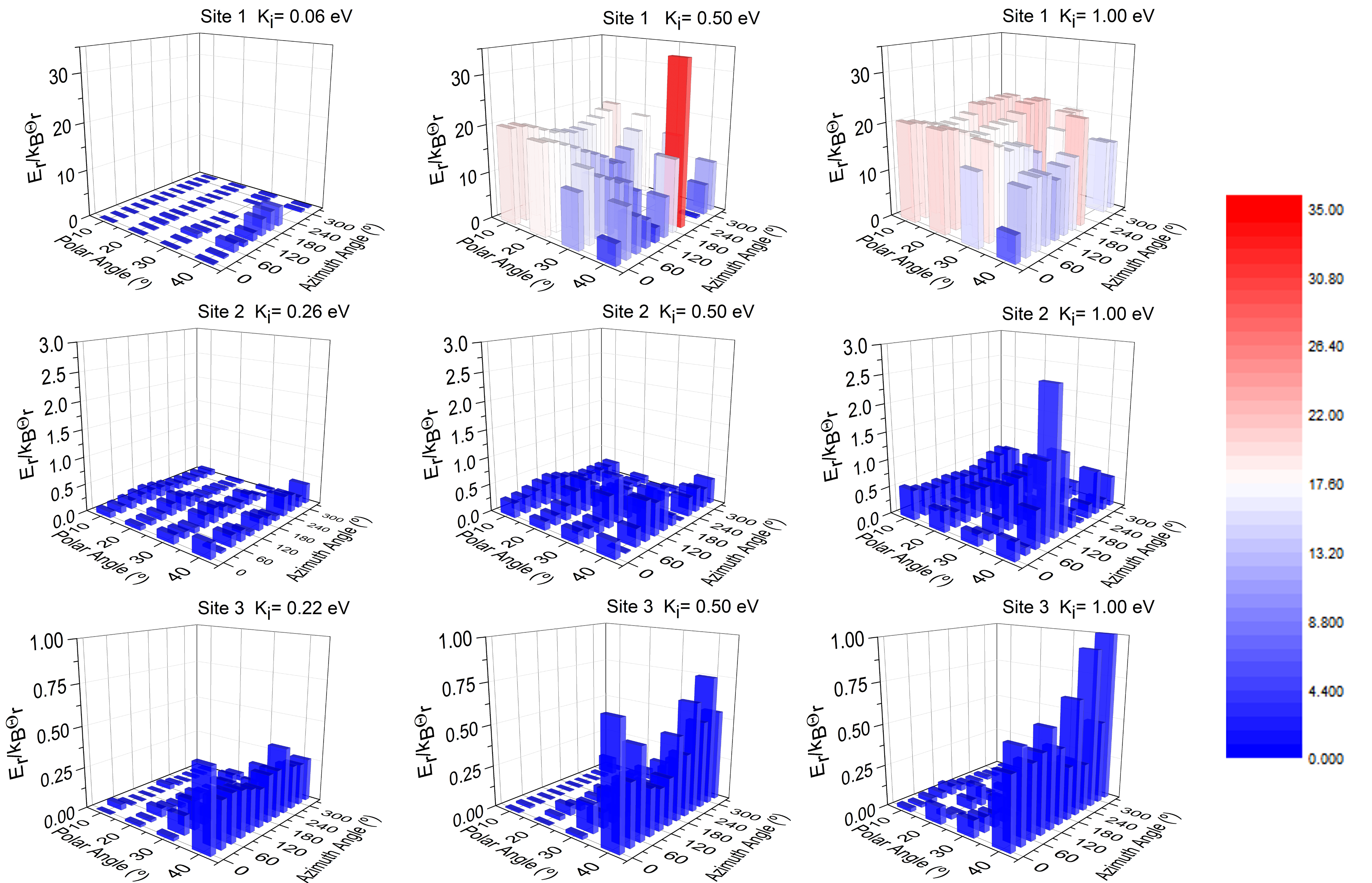}
    \caption{Ratio between the rotational energy of the newly formed $\mathrm{H}_{2}$ molecule and its characteristic energy of rotation ($k_{B}\,\Theta_{r}$).}
    \label{fig:Rotational}
\end{figure*}

The next step in the ER mechanism is the abstraction of the chemisorbed hydrogen atom by an incoming $\mathrm{H}$ atom, giving rise to the hydrogen molecule and recovering the coronene molecule, $\text{coronene-H} 
 +\mathrm{H}\cdot \rightarrow \text{coronene} + \mathrm{H_2}$. The abstraction happens through a radical recombination. Both the incoming hydrogen atom and the monohydrogenated coronene  are open-shell, while the reaction products are closed-shell in their ground states. Thus, abstraction can, in principle,  proceed through two different reaction channels: singlet and triplet electronic states. However, the singlet mechanism is more likely because it is barrierless \citep{2008_Rauls, Rasmussen_2011}  and quite exothermic (-2.7 to -3.5 eV, see Table \ref{tab:chem_energies}) while the triplet has an activation barrier ($\sim 0.4$\,eV) \citep{barrales1}. Hence, here we only focus on molecular dynamics on the singlet state. Although it is computationally cheaper to use a closed-shell wave function for the singlet, it results in an incorrect representation of the PES at large distances between hydrogens. Therefore, we will always use a wave function with broken spin-symmetry that allows us to recover the correct dissociation limit of $\mathrm{H}_2$.\\  

\begin{table}
\centering
	\caption{Reaction energy $\Delta\mathrm{E}^{\circ}$, impact energies  of hydrogen projectiles $\mathrm{K}_{i}$, average energy to dissipate $E_{D}$, and  the ratio between the rotational energy
of $\mathrm{H}_{2}$ and its characteristic energy of rotation $\langle \mathrm{E_{r}}/k_{B}\Theta_{r}\rangle$. All energies values in eV.\\
$E_{D} = \langle \Delta V\rangle + K_{i}$, where $\langle \Delta V\rangle$ is the average change in potential energy.}
	\label{tab:chem_energies}
	\begin{tabular}{ccccc} 
		\hline\hline
		Site $C_{i} (i = 1-3)$ & $\Delta\mathrm{E}^{\circ}$ & $\mathrm{K}_{i}$ & $E_{D}$ & $\langle \mathrm{E_{r}}/k_{B}\Theta_{r}\rangle$\\ 
		\hline\hline
		   & & 0.06  & 2.0 & 0.54 \\
		 1 & -2.7 & 0.50 & 2.4 & 12.50 \\
		   & & 1.00 & 3.0 & 17.77 \\\hline
		   & & 0.26 & 2.3 & 0.15 \\
		 2 & -3.4 & 0.50 & 2.5 & 0.30 \\
		   & & 1.00 & 3.0 & 0.50 \\\hline
		   & & 0.22 & 2.3 &  0.11 \\
		 3 & -3.5 & 0.50 & 2.6 & 0.16 \\
		   & & 1.00 & 3.0 & 0.18 \\
 		\hline\hline
	\end{tabular}
\end{table}

As in the previous case, the projectiles are initially $\SI{5}{\angstrom}$ away from the target atom (in this case the chemisorbed hydrogen atom) to prevents interactions between the surface and the projectile at the beginning of the molecular dynamics. The same impact energies were used as in the previous step. Our findings indicate that the abstraction of $\mathrm{H}_{2}$ is a highly probable process. In sites 2 and 3, regardless of the impact parameters,  all projectiles achieve the abstraction of the chemisorbed hydrogen atom and the formation of $\mathrm{H}_{2}$.  In site 1,  $\sim 92\%$ of the incoming projectiles successfully abstracted the chemisorbed hydrogen atom, while the remaining $\sim 8\%$ resulted in $\mathrm{H}$ projectiles with large polar angle ($\theta \geq 30^{\circ}$) that spill-over to a neighbor carbon,  leading to a doubly hydrogenated molecule. This phenomenon was also observed in previous research \citep{2008_Rauls}, which showed that the process is barrierless and aligns with our results.\\

Since $\mathrm{H}_{2}$ abstraction is an exothermic reaction, the coronene and the nascent $\mathrm{H}_{2}$ molecule must accommodate not only the kinetic energy of the incident H atom but also the energy released in the reaction (see Table \ref{tab:chem_energies}). Despite the coronene has enormous mass compared to H, it accommodates   about  $\sim 15\%$ of total excess of energy ($1-3$\,eV)  as vibrational energy, which corroborates the ability of coronene (PAHs in general) to act as a sink of energy for the $\mathrm{H}_2$ formation in the ISM. In a recent computational study, \citep{pantaleone2021} found that water ice surfaces absorb  $\sim 50\%$ of the energy released in the $\mathrm{H}_2$ formation. In their simulation H atoms carried no initial kinetic energy, which let time for a longer interaction between H and surface and and a more effective dissipation of energy.\\

The remaining  $\sim 85\%$ of the excess of energy is carried by the nascent hydrogen molecule. After its release, the $\mathrm{H}_{2}$ molecule acquires a large translational kinetic energy, which on average represents between 20 to 60\% ($\sim 0.4 - 1.25$\,eV) of the excess of energy, $E_D$. The translational  energy  of $\mathrm{H}_{2}$ molecules depends weakly on the polar angle of the orientation of incoming $\mathrm{H}\cdot$ projectile, as it can be seen in Figure \ref{fig:MTKE}. Two slight trends are observed, though.  $\mathrm{H}_{2}$ molecules released from the edge carbon ($C_1$) exhibit an increasing  of the translational  energy as the incoming projectile deviates from normal impact. Contrary, molecules released from inner carbons ($C_3$ and $C_3$) carry less translational energy as the impact deviates from normal.

\subsection{Rotational and vibrational states of the nascent $\mathrm{H}_{2}$}

The vibrational state ($\upsilon$) of the released $\mathrm{H}_{2}$ was approximated by matching the vibrational energy (measured $\sim$ 30 fs after $\mathrm{H}_{2}$ bond breaking with coronene) to closest eigenvalue of a Morse potential that fits $\mathrm{H}_{2}$ dissociation PES (see Section 3 in Supporting  Information). Figure \ref{f:H2_v} shows the fraction of molecules formed with a given $\upsilon$  that come from impacts at different polar angles. All  $\mathrm{H}_{2}$ molecules desorb in  vibrational excited states with $\upsilon$ as low as 3 and as large 10. These degree of vibrational excitation is align with other evidence. For instance, $\mathrm{H}_2$ formed on graphite surfaces occupies vibrational levels around $\upsilon=3-5$ \citep{casolo2013insights,latimer2008} and that formed on water surfaces occupy higher levels ($\upsilon=1-8$) \citep{Roser2003,yabushita2008}. Vibrational excitation depends on the impact energy and orientation of $\mathrm{H}\cdot$ projectiles. When $\mathrm{H}\cdot$ impacts on the H chemisorbed on the outermost carbon ($C_1$) at low angles ($40^{\circ}$), the released $\mathrm{H}_{2}$ occupies lower vibrational states ($\upsilon\geq 3$) with a broader distribution of $\upsilon$ ($3\leq \upsilon \leq 7$). However, for inner sites, the vibrational populations of the newly formed $\mathrm{H}_{2}$ molecules behave differently. When the projectile moves away from normal collision, the molecules tend to have a higher vibrational state (ranging in $7\leq \upsilon \leq 10$). As the impact energy increases, the occupation of vibrational states broadens. However, for the external site, there is a significant dependence on the impact angle. Molecules desorbing from the external carbon carry less vibrational energy than  those detached from internal carbons. This marked difference in vibrational energy is  traced to the different topologies of the PES around these sites.\\

The rotational energy ($E_{r}$) of the released $\mathrm{H}_{2}$ was calculated from the angular momentum at the turning points. Rotation is the degree of freedom that accommodates the smallest part of the excess energy, not exceeding 10\% in all cases. This a reasonable situation because the $\mathrm{H}_{2}$ molecule has a low moment of inertia, and hence a large rotational constant/temperature ($\Theta_r=85.3\,K$ \citep{McQuarrie_SM}). Our simulations show that in the low impact-energy regime, released molecules are far from achieving the thermodynamic limit of the distribution of rotational states ($E_{r}/k_{B}\Theta_{r}<1$), see Figure~\ref{fig:Rotational}. The dependence of rotational energy on impact energy is strongly dependent on where the abstraction takes place.  Even at collision energies as high as 1 eV, the molecules released from the inner carbons do not reach the thermodynamic limit of the rotational energy distribution. ($E_{r}/k_{B}\Theta_{r}<1$).  Instead, in the edge carbon all the molecules are released with $E_{r}>>k_{B}\Theta_{r}$. Note that even in the latter case, the rotational cooling capacity of the reaction site is limited and saturates with increasing impact energy (compare $C_1$ $K_i = 0.5$ eV and $K_i = 1.0$\,eV). \cite{yabushita2008} found that $\mathrm{H}_2$ products (in thermal equilibrium) of the photolysis of amorphous water ice H carry rotational energies in the range of $1-15\times 10^{-20}\,J$, which is equivalent to $E_{r}/k_{B}\Theta_{r}<15$. This is aligned with our result that rotational energy of $\mathrm{H}_2$ saturates at about   $E_{r}/k_{B}\Theta_{r}=15-20$ (see upper right panel in  Figure~\ref{fig:Rotational}).\\

\subsection{Implications of vibrationally excited $\mathrm{H}_{2}$ in the ISM}

Vibrationally excited $\mathrm{H}_{2}$ in space is primarily generated through different mechanisms and sources. Shock waves and far-ultraviolet fluorescence are two main sources, as indicated by \cite{Beckwith1978,sternberg1989excitation, Gatley1987,sheffer2009,Gnaci2013,rachford2014}. Additionally, X-ray photons and cosmic rays \citep{Padovani2022,Gaches2022A} have been suggested as potential sources, along with $\mathrm{H}_{2}$ formation itself \citep{wakelam2017h2}. For example, \cite{Chang2021} observed that over $90\%$ of the $\mathrm{H}_{2}$ formed from the UV photo-dissociation of $\mathrm{H}_{2}\mathrm{O}$ exhibited vibrational excitation at $\upsilon=3$. Similarly, \cite{Zhao2022} conducted a study on the fragmentation dynamics of   $\mathrm{H}_{2}\mathrm{S} + h\nu \longrightarrow \mathrm{S}(^{1}\mathrm{S}) + \mathrm{H}_{2}(\mathrm{X}^{1}\Sigma_{g}^{1})$  and found that $\mathrm{H}_{2}$ molecules formed with a wide range of vibrational states, with a peak at $\upsilon = 5$. Regarding reactions on surfaces, \cite{gough1996} discovered that $\mathrm{H}_{2}$ formed on graphite and other carbonaceous surfaces can populate high vibrational states, such as $\upsilon = 7$.
\\
Based on our simulations, we have observed that nascent hydrogen molecules desorbing from the edge sites of coronene, and likely from all polycyclic aromatic hydrocarbons (PAHs), exhibit excited vibrational states similar to those found in the experiments conducted by \cite{gough1996} and  \cite{Chang2021}. Interestingly, hydrogen molecules formed in internal sites of PAHs show even higher levels of vibrational excitation compared to those generated through photo-chemical reactions. On the other hand, 
the lifetime of vibrationally excited hydrogen molecules is on the order of days \citep{Chang2021, sheffer2009, 2000Draine}, mainly due to their Einstein A-coefficients, which range from approximately $10^{-7}$ to $10^{-6} s^{-1}$ \citep{Coppola2016, Wolniewicz1998}. This implies that $\mathrm{H}_{2}$ molecules formed on PAHs through the ER mechanism could serve as active reagents in the ISM chemistry. Vibrationally excited $\mathrm{H}_{2}$ molecules can participate in chemical reactions that are otherwise challenging to occur in the ground state, as their internal vibrational energy enables them to overcome activation barriers \citep{agundez2010chemistry}. This is particularly relevant for the initial stages of heavy element chemistry, involving the formation of hydrides through gas-phase reactions
\begin{ceqn}
 \begin{align}
           \mathrm{X} + \mathrm{H}_{2} & \longrightarrow \mathrm{XH} + \mathrm{H} 
\end{align}
\end{ceqn}
where $\mathrm{X}$ could be carbon, oxygen, nitrogen and sulfur atoms or their cations \citep{Stecher1972,Zanchet2019,2022AA_Goicoechea}.\\

\section{Conclusions}
We studied the energy dissipation in the $\mathrm{H}_{2}$ formation process via the Eley-Rideal mechanism onto a coronene surface by means of Born-Oppenheimer (\textit{ab initio}) Molecular Dynamics.  Our simulations show that for the  chemisorption  of hydrogen (first step of the Eley-Rideal mechanism),  the sticking probability depends on both the kinetic energy of the projectile and in its orientation (polar angle). The general tendency observed is that the sticking probability decreases as the projectile moves away from the normal orientation. Moreover, internal sites ($C_2$  and $C_3$, see Figure~\ref{fig:Cor_positions}) exhibit a lower sticking probability compared with the edge site ($C_1$). In contrast, it is the internal sites that are capable of dissipating a greater amount of the excess energy acquired due to chemisorption in a shorter time scale compared to the edge site.\\

For the second step of the Eley-Rideal mechanism, the abstraction and formation of $\mathrm{H}_{2}$, our simulations  show that the reaction is a highly probable process under the studied conditions. All incoming projectiles achieve abstraction at the inner sites and over $90\%$ at the edge site. About $\sim 15\%$ of the energy is dissipated to the coronene molecule as vibrational energy.  The remaining energy is transferred to the nascent hydrogen molecule. $\mathrm{H}_{2}$ detaches from  coronene  with a large amount of translational kinetic energy ($ \geq \SI{0.4}{eV}$) and in a highly excited vibrational state ($\upsilon \geq 3$). The activation of the rotational degrees of freedom is the mechanism that least contributes to the partition  of the excess energy linked to the reaction ($\leq 10\%$).\\

We also discussed how the formation of molecular hydrogen on PAHs through ER mechanism may represents a further source of vibrationally excited $\mathrm{H}_{2}$ observed in the ISM and play the role of activated reagent on the ISM chemistry.

\section*{Acknowledgements}

This research was funded by FONDECYT through projects 1220366, 1231487, 1220715 and by the Center for the Development of Nanosciences and Nanotechnology, CEDENNA AFB 220001.  NFB gratefully acknowledges to ANID by his national MSc fellowship year 2022 number 22220676. Powered@NLHPC: This research was partially supported by the supercomputing infrastructure of the NLHPC (ECM-02).

\section*{Data Availability}

All data necessary to reproduced and check this work (molecular structures and MD-trajectories) is available in the GitHub repository at \href{https://github.com/nfbarrera/Formation-of-H2-on-polycyclic-aromatic-hydrocarbons}{https://github.com/nfbarrera/Formation-of-H2-on-polycyclic-aromatic-hydrocarbons}.



\bibliographystyle{mnras}
\bibliography{Referencias} 



\section*{SUPPORTING INFORMATION}
Supplementary data are available at \textit{MNRAS} online.\\

\noindent
Optimized Cartesian Coordinates and Energies.\\
\textbf{Table S1.} Quantification of the conservation of energy for a set of AIMDs with DFTB and DFT.\\
\textbf{Figure S1.} Moving average of the change in potential energy for the dynamics calculated at the PBE/6-31G(d) and DFTB/MIO theoretical levels at site $\mathrm{C}_{1}$.\\
\textbf{Figure S2.} Moving average of the change in potential energy for the dynamics calculated at the PBE/6-31G(d) and DFTB/MIO theoretical levels at site $\mathrm{C}_{2}$.\\
\textbf{Figure S3.} Moving average of the change in potential energy for the dynamics calculated at the PBE/6-31G(d) and DFTB/MIO theoretical levels at site $\mathrm{C}_{3}$.\\
\textbf{Figure S4.} Morse potential and its eigenvalues for $\mathrm{H}_{2}$.






\bsp	
\label{lastpage}
\end{document}